# Origins of eukaryotic excitability

**Kirsty Y. Wan**[1,2] **and Gáspár Jékely**[1,3]

[1]Living Systems Institute, University of Exeter, EX4 4QD, UK
[2]College of Engineering, Mathematics, and Physical Sciences, University of Exeter, UK
[3]College of Environmental and Life Sciences, University of Exeter, UK

Email: k.y.wan2@exeter.ac.uk (KYW), g.jekely@exeter.ac.uk (GJ)

**Abstract:** All living cells interact dynamically with a constantly changing world. Eukaryotes in particular, evolved radically new ways to sense and react to their environment. These advances enabled new and more complex forms of cellular behavior in eukaryotes, including directional movement, active feeding, mating, or responses to predation. But what are the key events and innovations during eukaryogenesis that made all of this possible? Here we describe the ancestral repertoire of eukaryotic excitability and discuss five major cellular innovations that enabled its evolutionary origin. The innovations include a vastly expanded repertoire of ion channels, endomembranes as intracellular capacitors, a flexible plasma membrane, the emergence of cilia and pseudopodia, and the relocation of chemiosmotic ATP synthesis to mitochondria that liberated the plasma membrane for more complex electrical signaling involved in sensing and reacting. We conjecture that together with an increase in cell size, these new forms of excitability greatly amplified the degrees of freedom associated with cellular responses, allowing eukaryotes to vastly outperform prokaryotes in terms of both speed and accuracy. This comprehensive new perspective on the evolution of excitability enriches our view of eukaryogenesis and emphasizes behaviour and sensing as major contributors to the success of eukaryotes.

## Introduction

Cellular excitability is the capacity to generate highly nonlinear responses to stimuli, often over millisecond timescales. During eukaryogenesis, cells evolved new mechanisms of excitability to sense and react rapidly to their environment. In all cells, excitability is underpinned by the thermodynamics of interfaces. Interfaces are formed by biomembranes that bind regions with different ionic compositions. Excitability emerges as a biophysical consequence of charge separation across biological membranes. This is regulated by the passage of ions between different cellular compartments through ion channels or biochemical signals initiated by metabotropic receptors. These ionic currents then regulate effector systems including the cilium or contractility apparatus. The ionic homeostasis of the compartments is maintained by active pumping by ATPase pumps [1]. Biosensing is achieved whenever this homeostasis is disturbed from its equilibrium or steady state, by perturbative influences, which can then propagate rapidly and directionally across the membrane [2].

In general, eukaryotic cells display more complex behaviours than prokaryotes (archaea and bacteria). The differences often are not only quantitative but qualitative. During eukaryogenesis, radically new forms of sensing and reacting to stimuli have evolved. This enabled deterministic navigation in eukaryotes, particularly the direct tracking of gradients or the movement along vectorial cues in three dimensions in open water (3D taxis). Many eukaryotes also actively select particles during feeding, explore substrates, or undergo regulated cell-cell fusion during sex. Phagotrophy is another characteristic which distinguishes eukaryotes from other forms of life.

What are the evolutionary origins of these new forms of behaviour? Are there universal features that we can identify, which clearly set eukaryotes apart from prokaryotes? Here we attempt to trace the evolutionary origins of eukaryotic excitability by invoking a combination of cell biology and physical principles.

This paper is organised into two main sections. We begin with a comprehensive overview of eukaryote-signature behaviours and contrast this to those exhibited by prokaryotes. We proceed to argue that beyond changes in gene complements, several major cellular innovations were likely critical to the emergence of these new forms of behaviour. These all evolved during eukaryogenesis and were likely to have been present in the last eukaryotic common ancestor (LECA). Key structural innovations include changes in the complement of membrane channels, compartmentalisation, and new types of motility-generating appendages (eukaryotic cilia/flagella and pseudopodia). Furthermore, a general increase in cell size and compartmentalisation could have fundamentally changed the biophysical regimes accessible to eukaryotic cells. In this article, we present a new perspective of how these cellular innovations prompted the 'new physics' that may have unlocked the evolution of new sensory and response strategies that were previously unavailable to prokaryotes.

## 1. Forms of excitability in eukaryotes

In the first section, we give an overview of the forms of excitability, sensing and response, that are characteristic of eukaryotes, and contrast these strategies with those encountered in prokaryotes. We first focus on single-cell strategies for tactic navigation. Self-locomotion, or motility, is an important feat which not only increases the efficacy of environmental exploration, but also enables cells to move toward more favourable environments, and away from harm. Motility control in the presence of spatiotemporal gradients is an ideal testbed for evaluating a cell's sensory performance. Apparently very different microscopic sensorimotor rules can nonetheless lead to similar macroscopic or end outcomes, namely, some form of net migration towards or away from a stimulus.

Among motile organisms, strategies for navigation are often diverse and highly organism-specific. There are three major strategies for cells to track gradients of external cues (e.g. chemicals, light, temperature), which we shall refer to as stochastic navigation, spatial sensing and helical klinotaxis. This is inspired by the classification of Dusenbery [3], and also similar to the convention adopted by Alvarez *et al.* [4] in the context of chemotaxis. We shall seek to understand how the propensity for prokaryotes and eukaryotes to adopt different strategies may have arisen from adaptations to different physical regimes.

In addition, there are passive forms of orientation which we will not discuss in detail here. These include magnetotaxis in some proteobacteria and an euglenid alga [5–7]. There are further idiosyncratic forms of environmental tracking which do not fall into any of the above navigation categories, such as active regulation of buoyancy in non-motile diatoms in order to move up and down in the water column [8] [9].

With the advent of enhanced sensory and navigational capabilities, eukaryotes became capable of more sophisticated behavioural sequences (summarised in Figure 1). These include nonlinear responses to mechanical stimulation, and complex membrane dynamics, including cell engulfment and cell-cell fusion. Orderly sequences of excitable actions manifest themselves in the everyday processes in the life of a eukaryotic cell, for example chemosensing or tracking of other cells by chemotaxis, followed by engulfment (phagocytosis) or cell fusion (sex).

A. **Stochastic or non-oriented navigation**

The response of an organism or cell to stimuli is not always directional. Instead *stochastic navigation* strategies – also termed *kineses* – exist (Figure 1a). This form of navigation is most widespread among prokaryotes but also occurs in some eukaryotes. Stochastic navigation is a form of temporal sensing, during which changes in the concentration, e.g. of a chemical, are detected as the cell moves along a gradient. The most common strategy among prokaryotes consists of straight runs followed by brief turns [10] or tumbles to reorient the trajectory [11]. During positive chemotaxis, a sufficient rate of increase in the chemoattractant concentration will suppress tumbles, biasing the movement up the gradient [12]. Run-and-tumble chemotaxis does not involve directed turns and can therefore be referred to as statistical or indirect chemotaxis [3], or *chemokinesis*. Variations include run-and-reverse, run-reverse-and-flick [13], or run-and-stop [14]. It has been reported that some polarly flagellated bacteria exploit a buckling instability to steer, producing a peaked distribution around preferred turning angles [15]. These kinetic strategies depend on two-component signalling and generally apply to bacterial and archaeal chemotaxis and phototaxis [16,17]. Cells achieve biased migration according to generalised velocity-jump random walk processes [18].

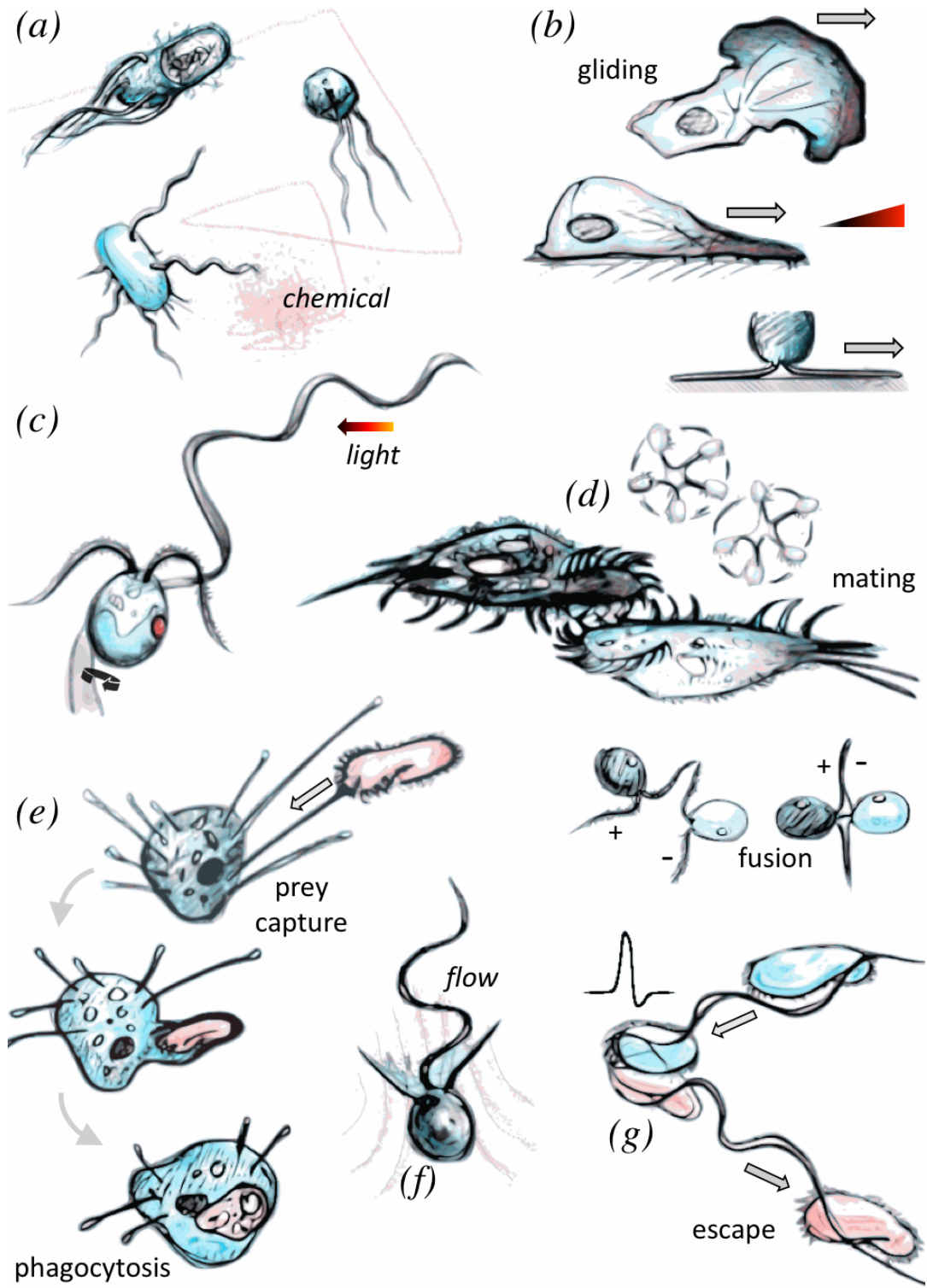

**Figure 1**. ***The many forms of cellular excitability.***
a) Stochastic or non-oriented navigation strategies (e.g. prokaryotic chemotaxis), b) sensing by spatial comparison (e.g. amoeboid chemotaxis), c) sensing by helical klinotaxis (e.g. flagellate phototaxis), d) cell-cell recognition as a prelude to fusion (e.g. ciliate conjugation and gametic fusion in *Chlamydomonas*), e) active feeding by selective engulfment of prey organisms, f) mechanosensitivity and flow interactions, g) ultrafast escape responses and reversal of ciliary beating by action potentials.

Stochastic trial-and-error navigation by temporal sensing also exists among eukaryotes but is not as well-characterised. Recently it was shown that aerotaxis (tracking of oxygen gradients) in choanoflagellates relies upon such a stochastic navigation strategy [19]. Many ciliates are capable of multiple gradient sensing strategies which may include a stochastic component, depending on the nature of the stimulus, or irritant. Bacterial chemoattractants can also stimulate chemokinesis in *Tetrahymena*, leading to an increase in swimming speed [20]. The capacity for chemoattractants (e.g. glutamate, folate) and chemorepellents (toxins) to induce positive and negative kinetic effects in ciliates including *Tetrahymena* and *Paramecium* is well-established [21,22]. Different chemical signals induce a combination of more subtle movement responses including klinokinesis (random reorientations) and orthokinesis (changes in speed). When exposed to extracellular GTP, some ciliates exhibit repetitive back-and-forth motion, a more complex form of excitability mediated by membrane potential [23].

### B. Deterministic steering by spatial comparison

In contrast to stochastic kineses, organisms can also tune their orientation with respect to a cue. Taxes are further subdivided [24] into *tropotaxis* when gradients are measured directly at two different locations at the same time (spatial sensing), usually requiring at least two spatially-separated receptors, and *klinotaxis* when gradients are measured sequentially at two different times (temporal sensing) at the same receptor, while the organism is in motion (see next section).

Sensing by spatial comparison (Figure 1b) is a form of navigation which is a eukaryotic characteristic, with a few exceptional prokaryotic cases. Spatial chemosensing or light sensing relies on the differential excitation of one part of a cell coupled to directional movement along a gradient or vector. Among eukaryotes, spatial chemosensing is widespread across many supergroups and possibly traces back to the LECA. Most amoeboid eukaryotes are thought to perform chemotaxis through spatial sensing [25].

Amoeboid cells migrate by directional chemotaxis towards bacteria or other food sources. The mechanism is spatial sensing coupled to the directional extension of pseudopodia [26]. Such amoeboid chemotaxis has been demonstrated in members of many eukaryotic clades. Amoeboid cells of the excavate amoeboflagellate *Naegleria fowleri* (Discoba, Heterolobosea) show chemotaxis towards bacteria [27]. Similar directed chemotactic migration is present in *Acanthamoeba* [28], *Entamoeba* [29], *Hartmannella* [30], *Dictyostelium*, and amoeboid cells in animals [26]. Directional chemotaxis also functions in syncicial forms like the plasmodia of the slime mold *Physarum* [31].

Chemotaxis in *Dictyostelium* is perhaps the best studied from a biophysical perspective [32]. In the absence of gradients, pseudopods extend randomly, but extensions become localised when gradients are detected. Cells can sense gradients of only 2% front-to-back, and can migrate towards travelling waves of chemoattractant [33]. In very shallow gradients, cells adopt a much more stochastic mechanism in which pseudopods extend randomly but form with a frequency which is chemoattractant-dependent [34]. *Dictyostelium* can also reconfigure its motility machinery to migrate against shear flows [35,36]. This active response relies on spatial sensing of differential hydrodynamic forces; just as during chemotaxis, directional pseudopodial motility occurs by an actin-polymerization-dependent mechanism.

There are other – likely more derived – forms of tropotaxes among eukaryotes. The euglenoid *Peranema* glides on surfaces and can efficiently move towards sources of food [37]. A long anterior flagellum and a short ventral flagellum are attached to an elongated cell body – which is itself capable of amoeboid-like movement in the absence of the flagella. Gliding by cilia is present in many species, and relies on a surface-mechanism mediated by interactions between transmembrane proteins and the substrate [38]. Red algae glide on surfaces and can move towards light by directional steering [39,40]. Since red algae generally lack cilia and helical swimming, their phototactic movement may rely on spatial sensing (shading or focusing by lipid droplets), and directional growth of cellular protrusions [41]. Filamentous eukaryotes with cell walls show directed orientation or growth along chemical gradients, including fungal hyphae [42], the pollen tube of plants [43], and plant roots undergoing hygrotropism [44].

Prokaryotes display spatial sensing only very sporadically or only as collectives. Multicellular communities of some bacteria display collective behaviour reminiscent of deterministic tracking by spatial comparison. Swarming *Pseudomonas aeruginosa* can track chemical trails by collective movement, even if the individual bacteria only move randomly [45]. Swarm colonies of the bacterium *Rhodospirillum centenum* show directional phototaxis through an unknown direction-sensing mechanism, possibly exhibiting colony-level shading [46,47].

At the single cell level, prokaryotes are too small for spatial sensing to be effective (Section 2F), with few exceptions. Some relatively large (2 x 6 μm) bipolar flagellated vibrioid bacteria can track steep oxygen gradients and perform directional turns, likely by spatial sensing [48]. The cells are able to detect oxygen gradients along a distance of several microns along their long axis. Turning may rely on the faster rotation of the flagellar bundle at the cell pole exposed to higher oxygen levels. Another unusual form of spatial sensing is found in the cyanobacterium *Synechocystis*, which is able to follow directional light cues. These cells act as spherical microlenses to focus incoming light to the opposite side of the plasma membrane. This localised stimulus induces motility in the direction of light [49].

### C. Deterministic steering by temporal comparison

Klinotaxis relies on temporal comparisons and chiral self-motion (Figure 1c). This is arguably the most sophisticated of cellular navigation strategies, and occurs almost exclusively in eukaryotes. Helical self-propulsion at low-Reynolds numbers arises naturally when cell-shape asymmetries are combined with periodic stroke patterns (e.g. of cilia and flagella) [50,51]. Direct taxes employing temporal sensing require the helical trajectories to have sufficiently large amplitude, as found in most freely swimming eukaryotes. During helical turns in a stimulus field, the cell tracks periodic changes in the stimulus, particularly in the direction perpendicular to the helix axis. By bending the helical trajectory in the stimulus direction, the cells can actively steer and migrate deterministically. Thus, helical klinotaxis is fundamentally different from stochastic navigation [3,4], and generally both more efficient and more robust to noise than other navigation types.

Diverse eukaryotes from distinct phyla use helical klinotaxis to track chemical gradients (such as diffusing from a food source) [52]. Helical klinotaxis has been described in various ciliates, the gametes of brown and green algae [53,54], the nonphotosynthetic green alga *Polytomella magna* [55], animal sperm [56], fungal zoospores [57], the heterotrophic flagellates *Cafeteria* [52] and *Euglena* [58], and the dinoflagellate *Peridiniopsis berolinensis* [59]. The gametes of many organisms are also guided by helical chemotaxis for external fertilization [60].

A special form of chemotaxis is a prelude to gametic fusion. The helical klinotaxis of flagellated male gametes towards female gametes is common. Since cilia [61], meiosis [62], and thus gametogenesis trace back to LECA, it is possible that the gametes of LECA found each other by helical klinotaxis before sex. In order for gametes to interact, they are often attracted to each other by pheromones – likely some readily diffusible substance. This attraction has been extensively studied in species of brown algae. When the female gametes of brown algae settle on a surface, they secrete sexual pheromones (olefinic C11-hydrocarbons) [63,64] which attract the much more motile, biflagellated male gametes. The male gametes can either swim freely in 3D helical trajectories, or swim close to the surface, moving in 2D circular paths (referred to as thigmotaxis). The chemoattractant pheromones influence the beat pattern of the flagella, increasing the curvature of the helical swimming path [65]. Helical klinotaxis is also adopted by monoflagellated gametes including fungal zoospores and animal sperm. The male gametes of the aquatic fungus *Allomyces* swim in a helical path interrupted by 'jerks'. The female gametes show little motility and secrete a pheromone (sirenin) which influences the swimming trajectories of male gametes, in order to guide their directional swimming [57].

In other examples of gametic attraction it is not yet clear whether the mechanism involves helical klinotaxis or stochastic chemokinesis. When both partners are motile, chemoattraction need not be unidirectional. In the ciliate *Blepharisma japonicum* [66] single cells swim more slowly with increasingly circular trajectories around higher concentrations of gamone. Each partner can secrete a different pheromone, which has the potential to attract cells of the opposite mating type. In *Spirostomum* and *Euplotes*, characteristic pre-conjugation rituals ('courtship' or even mating dances) have been described [67,68].

While chemical or thermal gradients are diffuse, other classes of physical stimuli (such as light, gravity, flows) are inherently directional. Phototaxis is a particularly striking example of helical klinotaxis [69]. It likely evolved at least seven times independently during eukaryotic evolution and may not have been a property of the LECA. However, as a biophysical strategy it is uniquely eukaryotic and was enabled by eukaryotic cellular organisation, excitability and movement pattern.

For example, the biflagellate alga *Chlamydomonas* has a single eyespot which leads to an eight-fold separation in perceived intensity from one side compared to the other [70]. A non-planar beat pattern leads to a helical trajectory which allows the organism to continuously scan 3D space and modulate the helix axis towards or away from the light. A similar strategy of helical phototaxis can be found in other protists, including *Euglena*, brown algal swarmers, and chytrid zoospores (reviewed in [69]).

In the colonial alga *Volvox carteri*, movement of thousands of flagella – which operate as individuals – achieve colony level phototaxis by virtue of their common stimulus-response function and positional distribution on a spherical body [71]. In *Euglena*, the interplay between eyespot sensor placement and an intensity-dependent reorientation strategy leads to more complex responses to patterned light [72,73].

The most extreme examples of light sensing in a single eukaryotic cell are found in certain predatory dinoflagellates which have large and highly unusual 'camera eyes', called ocelloids – built from endosymbiotically acquired components. This unique organelle has superior light-gathering optics and can refract light to a retina-like structure [74,75]. In the warnowiid dinoflagellate *Erythropsidinium*, the design and sophistication of its eye suggests that they can do more than sense light gradients (as in most other protist photoreceptors) [76]. Ocelloid eyes are even capable of an active rolling or pivoting motion [77]. It is suggested that it can detect circularly-polarised light – a tell-tale sign of a prey dinoflagellates (polarotaxis). This extraordinary sensory capacity coincides with an arsenal of cellular weapons (e.g. harpoons, nematocysts) which this organism uses to impale its prey. The klinotaxis mechanism in this case must be highly refined, though the connection between signal and action on the flagellar beat pattern is unclear.

In contrast to its widespread use in eukaryotes, klinotaxis is largely absent from prokaryotes. Perhaps the only example of true taxis in bacteria occurs in the large sulfide-oxidizing proteobacterium *Thiovulum majus*. *Thiovulum* cells are unusually large for a bacterium (5-25 μm) [78] and can swim at a speed of up to 615 μm/sec, one of the fastest swimming ever recorded for a bacterium [79]. *Thiovulum* cells swim with multiple flagella in a left-handed helix and can directly track a gradient of oxygen by bending the helical trajectory as a function of changes in concentration [80]. Large groups of cells also form large-scale fronts due to their chemotactic behaviour [81].

### D. Cell-cell recognition as a prelude to fusion

Sexual cycles consisting of meiosis and complete cell-cell fusion are unique to eukaryotes. From the perspective of excitability, the key properties of sex are the specific attraction and recognition by gametes, their complete fusion and the prevention of multiple rounds of fertilization. Eukaryotes undergo complete cell-cell fusion during sex and orchestrate this process with remarkable precision.

Mating starts with chemoattraction, often mediated by pheromones and helical klinotaxis (see above). Adhesion and fusion are regulated by genetically determined mating types [82]. In the ciliate *Euplotes patella*, which express 6 mating types and at least 3 different pheromones (gamones), this attraction is thought to be combinatorial [83]. Gamete recognition can be accompanied by dramatic changes in behaviour. In *Paramecium*, there is a marked reduction in swimming speed following recognition [84]. In the biflagellate alga *Chlamydomonas*, cells of opposite mating type collide randomly, before fusing to form quadriflagellate zygotes. Adhesion occurs via different mating-type-specific glycoproteins expressed on the flagella (agglutinins) [85].

Cilia and flagella often participate in cell-cell fusion as mating organelles (Figure 1d). In *Paramecium aurelia*, conjugation begins with adhesion of specific cilia, and cells become adhered at the anterior-ventral side [86]. Membrane fusion then occurs, at surfaces where cilia degenerate, and pores form to provide cytoplasmic bridges between the partners. In *Chlamydomonas*, isogamous gametes of opposite mating types adhere head-to-head along the length of their flagella, with the eyespots on the same side of the cell [87]. The plus mating type then extends a tubular, actin-based mating structure which fuses with a smaller structure in the minus mating type. The temporary quadriflagellate dikaryons are also phototactic [88]. In *Chlamydomonas*, flagellar adhesion is also light-sensitive, in fact light switchable [89], which may explain the necessity of light for gametogenesis in this organism [90]. In oogamous brown algae, the male gametes attach to the female gamete with their anterior flagellum [91]. After one gamete fuses, the others are excluded. This is ensured by the formation of a fertilisation barrier by the fusion of Golgi-derived vesicles underlying the plasma membrane [91]. It is unclear if initial chemotaxis and subsequent agglutinative contact are separate processes.

Among prokaryotes, cell-cell fusion is rare, and where it exists it is incomplete and reversible. In some haloarchaea, the exchange of genetic material can occur through incomplete cell-cell fusion, during which cells are connected by cytoplasmic bridges [92]. The selectivity of the mating process is regulated by surface glycosylation in *Haloferax volcanii* [93]. Among bacteria, transient outer membrane fusion has been observed in the myxobacterium *Myxococcus xanthus* [94]. In the spirochete *Borellia*, a study reported frequent outer membrane fusion and occasional inner membrane fusion [95].

### E. Active feeding by selective engulfment

Another key trait of eukaryotes is predation by whole-cell engulfment, which is often proposed as one of the main innovations driving eukaryogenesis [96,97], an idea not without its critics [98]. Particles, or even whole cells primed for engulfment may need to be first brought to a food vacuole by interception, track-and-capture, or by active filtration of self-generated currents [99].

Chemoattractants function to draw predators to prey, followed by specific and ordered activity sequences prior to prey engulfment, which may include cell attachment and the formation of specialised cellular protrusions. In the slime mold *Dictyostelium*, a GPCR, folic acid receptor 1, recognises both diffusible chemoattractants and surface molecules of bacterial prey [100,101]. The dinoflagellate *Oxyrrhis marina* uses a mannose-binding lectin as a feeding receptor for recognizing prey [102].

In many cases, phagotrophic predators are highly selective feeders. The heliozoan *Actinophry sol* can intercept and consume ciliate prey as large as itself by adhering to the ciliate with pseudopod extensions called axopodia, which then wrap around to completely enclose the prey [103]. Many protists possess specialised structures to facilitate prey immobilization and capture [104,105]. Phagotrophic euglenids actively shovel and manipulate prey organisms toward feeding grooves. The freshwater eukaryovorous euglenid *Heteronema* has been described to ensnare *Chlamydomonas* cells whole, by coordinating the action of multiple hook-bearing and mucus-covered flagella into the flagellar pocket [106]. The euglenid *Peranema* uses feeding rods to first pierce prey cells [107], before sucking out their contents directly into a feeding vacuole (*myzocytosis*). Photoautotrophic or osmotrophic species have much more reduced feeding apparatuses. Some dinoflagellates make use of a pseudopodial pallium to envelop and engulf large and awkwardly-shaped prey (e.g. diatoms) [108].

Most ciliates, despite having an ordered and largely rigid structure (pellicle), restrict phagocytosis to a single expandable feeding cavity. This structure, known as the cytostome, is often decorated with a ring of specialised cilia. After first paralyzing its prey with toxicysts, *Didinium* can engulf and pass a *Paramecium* cell as large as itself through its cytostome [109,110]. This process is associated with calcium-dependent electrical activity [111]. Hypostome ciliates make use of a cylindrical cytopharyngeal basket, which constricts to pass its filamentous prey (e.g. cyanobacteria) into a coil deep within the cytoplasm, at rates of up to 15 µm/s [112]. This is accompanied by rapid incorporation into the developing food vacuole, of new membrane recycled from vesicular fusion. Generally, the feeding cavity must enlarge and acidify before the prey can be digested [113].

In filter feeders, ensembles of cilia and flagella coordinate to pump fluid at relatively high speeds (up to 1 mm/s), often sweeping particles directionally into a feeding apparatus [114]. In collared choanoflagellates and the sessile ciliate *Vorticella*, food particles are sorted by size. In other species, mechanical filtration may be supplemented by adhesive surfaces for added particle selectivity.

Predatory behaviour also exists in prokaryotes [115], but due to their inability to engulf, proceeds in a very different way. Predatory bacteria generally inflict chemical lysis upon prey cells, or attack collectively. Single-cell strategies are rare. *Bdellovibrio* hunts bacteria [116] but the tracking mechanism does not appear to be chemosensory but rather dependent on hydrodynamic entrapment [117]. The unusual Planctomycete *Candidatus* Uab amorphum [118] can engulf other bacteria and picoeukaryotes, via a mechanism which resembles but is not thought to be homologous to eukaryotic phagocytosis. However this is a singular system which is not distributed among prokaryotes.

### F. Mechanosensitivity and flow interactions

Eukaryotes are particularly susceptible to mechanical stimuli and changes in membrane geometry. Many eukaryotes exhibit mechanosensitivity. This allows them to respond actively to hydromechanical signals transmitted remotely through the fluid, without need for direct contact with a potential predator or prey [119–121]. Marine ciliates can perform powerful jumps in response to predator-induced feeding currents at shear rates (magnitude of flow velocity gradients) of $1\sim 10$ s$^{-1}$ [122,123].

More graded responses to mechanical cues are controlled by specialised ion channels and transmitted directly through the membrane [124]. These may be localised to cilia and flagella, which display an active load response [125,126]. Hair-like extensions of cilia (mastigonemes) are also implicated in mechanosensing in *Chlamydomonas* [127]. A combination of rheosensing and chemosensing guides sperm motility through the mammalian oviduct. *Dictyostelium* reorients actively to shear flows [35].

The mechanosense is also manifest in gravitaxis, as well as gyrotaxis – the interaction between motility, gravity, and hydrodynamic torques in many marine protists and larvae. In the ciliate *Loxodes*, active control of membrane potential and statocyst-like organelles is thought to determine the sign of gravitaxis [128,129]. Some protists have also adapted to life in the ocean by exploiting sharp vertical gradients and undergoing ballistic diel migration [130]. However the gravity-sensing mechanism remains unclear, but is likely to involve a passive shape-dependent mechanical component [131], in addition to active regulatory mechanisms.

Planktonic microorganisms show a great deal of resilience against turbulent ecosystems [132]. For this they must be able to integrate and respond to multisensory information: such as light, chemicals, flows, gravity. Such single-cell responses can lead to large-scale population structure such as algal blooms, even formation of photo-gyro-gravitactic bioconvection patterns and instabilities [133,134]. In general, behavioural transitions are mediated by stimuli-dependent ionic currents and an excitable membrane, coupled to some form of self-locomotion. Cross-responses are even possible (where one type of stimulus elicits a change in the cell's response to another stimulus type), the male gametes of brown algae switch the sign of phototaxis from positive to negative when exposed to chemoattractants [135]. Self-movement in a fluid will alter local gradients, providing an opportunity for reafferent feedback.

Prokaryotes may be too small to respond actively to shear, i.e. gradients in flow velocity. Bacterial rheotaxis (reorientation to shear flow) has been shown to be a passive phenomenon [136]. However bacteria are capable of osmoregulation, and do possess ion channels which sense membrane tension [137], and mechanosensation in *E. coli* has been demonstrated to rely on voltage-induced calcium flux [138]. The torque produced by the flagellar motor (therefore rotation speed) is also load-dependent [139,140]. Other appendages (e.g. pili) for surface movement are candidate mechanosensors, and may reflect adaptation to substrate adherence for community-living and biofilm formation [141].

### G. Escape responses and action potentials

Stimuli which have the potential to harm or kill demand more immediate detection. This is fundamentally distinct from navigation or exploration, in terms of the timescales available for response. Most motile species harbour a form of phobic or emergency response distinct from its steady-state locomotion. Escape reactions are not strictly oriented – but commonly involve backward movement, sometimes with a negatively geotactic component [142]. Additional chemical self-defence strategies may be deployed, for example extrusion of trichocysts by *Paramecium* [143], or ejection of extrusomes in other ciliates [144].

In flagellate algae, abrupt changes in light intensity or intense photic stimuli induce rapid flagellar reversal and transient backward swimming [76,145]. In green algae this action may be mediated by the contractile root fibre which alters the angle between basal bodies [146]. Cells can also react at speed to unexpected mechanical stimuli. All-or-none contractions in the stalked ciliate *Vorticella* can occur at rates of 8 cm/s [147]. In some species of heliozoa, axopods can completely retract within 20 ms in order to draw in trapped prey for phagocytosis [147,148].

These fast reactions are usually induced by action potentials – unidirectional electrical pulses involving fast, regenerative changes in membrane potential. While all cells display some electrical activity, phylogenetic evidence suggests that the capacity to propagate action potentials may have been an ancestral eukaryotic trait supported by the LECA. These may have emerged in response to accidental membrane damage and sudden calcium influx [149]. Bioelectrical signalling in the form of action potentials occurs orders of magnitude faster than any other signalling modalities, e.g. chemical diffusion, protein phosphorylation etc.

In order to initiate fast escape responses, these may have been coupled directly to the motility apparatus – particularly to flexible, membrane-contiguous structures such as cilia and pseudopodia. Loss of voltage-gated sodium/calcium channels is further correlated with loss of cilia in many taxa. In protists, all-or-none action potentials occur almost exclusively in association with ciliary membranes [150–152], with the exception of some non-ciliated diatoms [153,154]. Graded potentials occur in amoeba, also for movement control [155].

In *Chlamydomonas*, action-potential like flagellar currents induce photophobic responses and flagella reversal (via voltage-gated calcium channels Cav2), while photoreceptor currents elicited much milder responses [156]. Here, mechanosensory channels (TRP11) are localised to the ciliary base, while Cav2 only to the distal regions of cilia [157,158]. In *Paramecium*, hyperpolarizations increase ciliary beat frequency, while depolarizations have the opposite effect and eventually lead to ciliary reversal. Depolarizations above a certain threshold result in action potentials, due to opening of Cav channels located exclusively in the ciliary membrane [159,160]. Potassium channels – also residing in the membrane – help restore the resting membrane potential.

Eukaryotes manipulate their membrane potential to achieve transitions between different behaviours. Complex bioelectric sequences have been recorded in association with integrated feeding and predation behaviours in *Favella* [161]. Repetitive behaviours arise from rhythmic spiking. In ciliates, rhythmic depolarisations control fast and slow walking by cirri [162], enable escape from dead ends [163], and courtship rituals in conjugating gametes [68,83]. In *Stentor*, action potentials produce whole-body contractions [164]. Finally, excitable systems operating close to bifurcations may admit limit cycles, which manifest as repetitive or rhythmic electrical spiking, and repetitive behaviours. Ultimately, this may lead to habituation [165,166].

In prokaryotes, action potential-like phenomena have been observed in biofilms [167] and also single cells. The archaeon *Halobacterium salinarium* shows a photophobic response characterised by a 180° reversal of its swimming direction induced by a reversal in the direction of flagellar rotation. At least some aspects of this response are likely mediated by changes in membrane potential by bacteriorhodopsin, a light-driven proton pump [168]. Action potential-like phenomena in prokaryotes are dissimilar from classical eukaryotic action potentials. The former is less reproducible, slower, and exhibits a broader distribution in pulse amplitude and duration [138].

| Forms of cellular excitability | Prokaryotic examples | Eukaryotic examples | Made possible by |
|---|---|---|---|
| **Stochastic navigation** | Bacterial and archaeal chemotaxis, archaeal phototaxis | Choanoflagellate aerotaxis, chemokinesis in some ciliates and flagellates | any moving appendage or motility mechanism |
| **Spatial sensing** | *Synechocystis*, some vibrioid bacteria | Amoebae, ciliates | spatially located sensor, large cell size |
| **Temporal taxes** | *Thiovulum* | Helical photo- and chemotaxis across eukaryotes | helical/chiral self-motion, fine motor control over cilia or flagella, temporal sensing, memory |
| **Cell fusion** | Some haloarchaea (incomplete), *Borellia* (mostly OM, incomplete) | All gametic fusion events | Cell-cell recognition, adhesion, in most eukaryotes mediated by cilia/flagella |
| **Active feeding by engulfment** | Planctomycete | Many eukaryotic phagotrophs | deformable membrane, cell recognition, sometimes by specialised appendages, internal digestion |
| **Mechanosensitivity and flow interactions** | osmosensation | Many eukaryotes | mechanosensory channels (e.g. TRP), membrane fluidity |
| **Escape responses and action potentials** | cable bacteria, some biofilms | Many eukaryotes | Voltage and calcium channels (e.g. Cav, Nav), often localised to cilia/flagella |

*Table 1. Forms of cellular excitability in eukaryotes are contrasted with those in prokaryotes.*
See main text for references.

## 2. Cellular and biophysical innovations underpinning eukaryotic excitability

In this part we give an overview of the cellular innovations which contributed to the emergence of new forms of excitability during eukaryogenesis. These are respectively, a) an extended repertoire of membrane receptors, channels and pumps, b) new forms of motility, c) endomembranes and mitochondria as ionic compartments and intracellular capacitors, d) a flexible and reconfigurable membrane, e) a larger size, f) new strategies for sensing. We identified these features as of major significance for the origin of eukaryotic excitability, but the list may not be exhaustive. We discuss how these factors contributed to the novel forms of excitability highlighted in the preceding sections and how they underpin new regimes of cellular biophysics that are only accessible by eukaryotic cells.

We shall embed our discussions within the most generally accepted framework for eukaryogenesis. This starts with an archaeal host related to the Asgard archaeal lineage [169]. This host acquired the mitochondrial symbiont – related to alphaproteobacteria – by internalisation. There are several versions of this model, and from our perspective some of the most intensely-debated details are less relevant (e.g. how early or late and through which intermediate stages did mitochondria evolve) [170]. We focus here only on the key eukaryotic novelties of membrane topology, motility, ionic currents and other ingredients necessary for excitability, and how these may be contrasted with prokaryotic cell biology. Of note, there are also alternative – and in our view less plausible – cell evolution scenarios for eukaryogenesis, e.g. involving three symbiotic partners [171], which we will not consider here.

### A. An expanded repertoire of ion channels, pumps and membrane receptors

In eukaryotes, there is a vastly expanded repertoire of membrane channels, pumps and receptors, distributed across a highly compartmentalized cell. Comparative genomics indicates much of this diversity evolved during eukaryogenesis in stem eukaryotes and was present in the LECA (e.g. [172]).

The various fast and slow ionic currents generated by receptors and ion-selective channels underlie the responses of eukaryotic cells to diverse sensory stimuli and injury. The regulation of motility, contractility, mechanosensation, tactic or temperature responses all rely on membrane excitability. The complexification and diversification of ion channels and receptor pathways was one of the major innovations that underpinned the evolution of the new forms of excitability in eukaryotes. In parasites, this diversity can be dramatically reduced. In parasitic trypanosomes, voltage-gated channels are represented by one type, a reduction from ten types in their free-living relative, *Bodo saltans* [173].

The regulation of calcium signalling illustrates the exuberance of systems eukaryotes evolved to control the flux of a single ion. Its influx and extrusion are regulated by various types of $Ca^{2+}$ channels and pumps, including store-operated $Ca^{2+}$ channels [174], $Ca^{2+}$ ATPases, voltage-gated and ligand-gated $Ca^{2+}$ channels, and transient receptor potential (TRP) channels.

The levels of free calcium are low in the cytoplasm and high in the ER. Intracellular $Ca^{2+}$ is kept low by the action of the PMCA (plasma membrane calcium-transporting ATPase), which counters the influx of $Ca^{2+}$ at the plasma membrane. The influx of $Ca^{2+}$ into the ER in turn is controlled by the SERCA (sarcoplasmic/endoplasmic reticulum calcium ATPase) $Ca^{2+}$ pumps. The ER and PM calcium systems are interlinked. The activation of PM receptors (e.g. some G-protein coupled receptors) through second messengers can gate the inositol trisphosphate receptor (InsP3R) in the ER to induce $Ca^{2+}$ release. The levels of $Ca^{2+}$ in the ER then influence store-operated $Ca^{2+}$ entry by plasma membrane $Ca^{2+}$ release-activated $Ca^{2+}$ (CRAC) channels (with the ORAI pore subunit [175]). Mitochondria are also involved in $Ca^{2+}$ signaling. Many eukaryotes harbour a mitochondrial $Ca^{2+}$ uniporter [176].

The core $Ca^{2+}$ transport systems of ER and PM channels and pumps, including SERCA, PMCA, IP3R, TRIC and ORAI, have homologs across diverse eukaryotes and were likely present in the LECA [172,177,178]. Some of the machinery of calcium signalling evolved in the common ancestor of apusozoa, animals and fungi representing a post-LECA diversification of ionic signalling (e.g. the sperm-specific CatSper $Ca^{2+}$ channel complex) [179].

Calcium signalling mediates diverse excitable phenomena across eukaryotic taxa. For example, in the ciliate *Tetrahymena*, phagocytosis depends on calcium signaling [180]. In fungi, pulsatile calcium signaling events accompany cellular events, including cell-cell contact and polarized growth [181]. In the slime mold *Dictyostelium*, speed modulation in migrating cells in response to shear stress is dependent on G-protein signalling acting through plasma membrane calcium channels and IP3-mediated internal calcium-store release [26,100,182]. In ciliates, voltage gated $Ca^{2+}$ channels regulate motility [150] and temperature responses [183].

Besides $Ca^{2+}$, several other ions are involved in excitability phenomena and are regulated by dedicated pathways. The *Paramecium* membrane for example has at least six different ionic conductances [184]. Diatoms and coccolithophores display voltage-activated cation and anion currents [185] [154]

Sensory receptor pathways responsible for mediating responses to external cues also originated early in eukaryotes and some were likely present in LECA. These include members of the transient-receptor potential (TRP) and G-protein coupled receptor (GPCR) families. TRP channels at the plasma membrane and in the ciliary membrane mediate several excitable phenomena. For example, TRP channels are involved in mechanosensation-induced bioluminescence in the dinoflagellate *Lingulodinium polyedra* [186]. Rheotaxis in shear-flow is dependent on a PKD2-like TRP channel in the amoeba *Dictyostelium discoideum* [187]. In the excavate *Euglena gracilis*, $Ca^{2+}$ influx via a mechanosensitive transient receptor potential (TRP) channel mediates negative gravitaxis [188]. In the green alga *Chlamydomonas reinhardtii*, there are several TRP channels involved in various sensory processes. The mechanosensory avoidance reaction of the cells requires the expression of TRP11, a TRPV-like ion channel, localised to the base of flagella [157]. Another flagellar TRP channel, CrPKD2, is important for flagella-dependent mating [189] whereas TRP1 is a thermo-sensitive channel that opens at increased temperatures [190].

Prokaryotes also have channels and transporters involved in various aspects of signalling and cell physiology such as chemotaxis [191–195]. These channels can be homologous to the eukaryotic channels, for example the trimeric intracellular cation-specific (TRIC) channels are conserved across bacteria, archaea and eukaryotes [196]. However, the prokaryotic channels are often simpler and sporadically distributed. For example, most eukaryotic voltage-gated channels are tetrameric whereas prokaryotic channels are monomers [197,198](but see [154]). The mitochondrial $Ca^{2+}$ uniporter is widespread in eukaryotes but found only sporadically in bacteria [176]. Eukaryotes thus distinguish themselves by the diversity and complexity of channels, pumps and receptors.

### B. New forms of motility: establishing fine control of cilia and pseudopodia

There may be as many as 18 distinct motility types across all forms of life [199]. Among these, notable eukaryotic motilities include free-swimming by cilia and migration by pseudopodia. The earliest eukaryote may have been an amoeboflagellate alternating between cilia-driven and amoeboid locomotion. The amoeboid-to-flagellate transition is a common lifecycle strategy among eukaryotes, occurring in rhizaria [200], amoebozoa, Naegleria [201], several opisthokonts including choanoflagellates [202], filastereans [203] and early-branching fungi [204].

These new forms of motility rely on eukaryotic-signature cell biology including the presence of a dynamic actin and microtubule cytoskeleton and membrane-bound cilia. Here we focus on the uniquely eukaryotic biology of cilia (confusingly, also referred to as flagella in their longer forms) and discuss how they might contribute to behaviour and sensing.

We briefly contrast the elaborate regulation, stroke patterns and gait control available for cilia-based motility, with the control of prokaryotic flagella and archaella [207–210]. All three appendage types share some physical similarities. They are all slender structures, which create drag anisotropy when moving through a fluid. Net propulsion is achieved by breaking time-reversal symmetry with propagating waves or chiral rotation. Asymmetric interactions also mediate gliding and crawling [211–213].

Beyond these similarities, cilia stand out with a unique propulsion-generating machinery that is very different from that of bacterial flagella or archaella. Bacterial flagella and archaella are extracellular structures, composed only of a few proteins plus a rotary motor and membrane-embedded base structure, whereas membrane-bound cilia have over 500 proteins [214]. Unlike either of the prokaryotic structures, which are driven by rotary motors from one end, dynein motors populate the entire length of cilia. This is known as distributed force actuation [215], in stark contrast with boundary actuation (from only one end) in the prokaryotic appendages.

The continuity of ciliary membranes as an extension of the plasma membrane enables fast reactions. Active amplification at these locations reduces transduction times from sensor to motor actuators (locomotor appendages), thereby leaving more time for sensory integration or sampling (see section F). At excitable interfaces, spontaneous fluctuations (due to noisy ion channels) of the resting membrane potential can amplify into action potentials. In flagellates and ciliates, this appears as spontaneous swimming reversals, which can occur in the absence of stimuli [206,216]. Artificial deciliation abolished the calcium-dependent excitability in *Paramecium* [217,218], further highlighting the possible ciliary origin of eukaryotic action potentials [219]. Leveraging mutants defective in ion-gating, several phospholipids were shown to be localised exclusively to the ciliary membrane of *Paramecium* [220].

While prokaryotes can manipulate the sense (and sometimes speed) of flagella rotation, the flexible and distributed architecture of cilia enables many more degrees of freedom. Dynein activity can be modulated intracellularly to produce a spectrum of beating modes within the same structure [221]. For instance, in green algal cilia, calcium induces a switch from a forward swimming mode with an asymmetric waveform (low-frequency) to a reverse mode with a symmetric waveform (high-frequency) in tens of milliseconds [206,222]. The beat pattern and asymmetry of sperm flagella is controlled by calcium and cAMP [223,224].

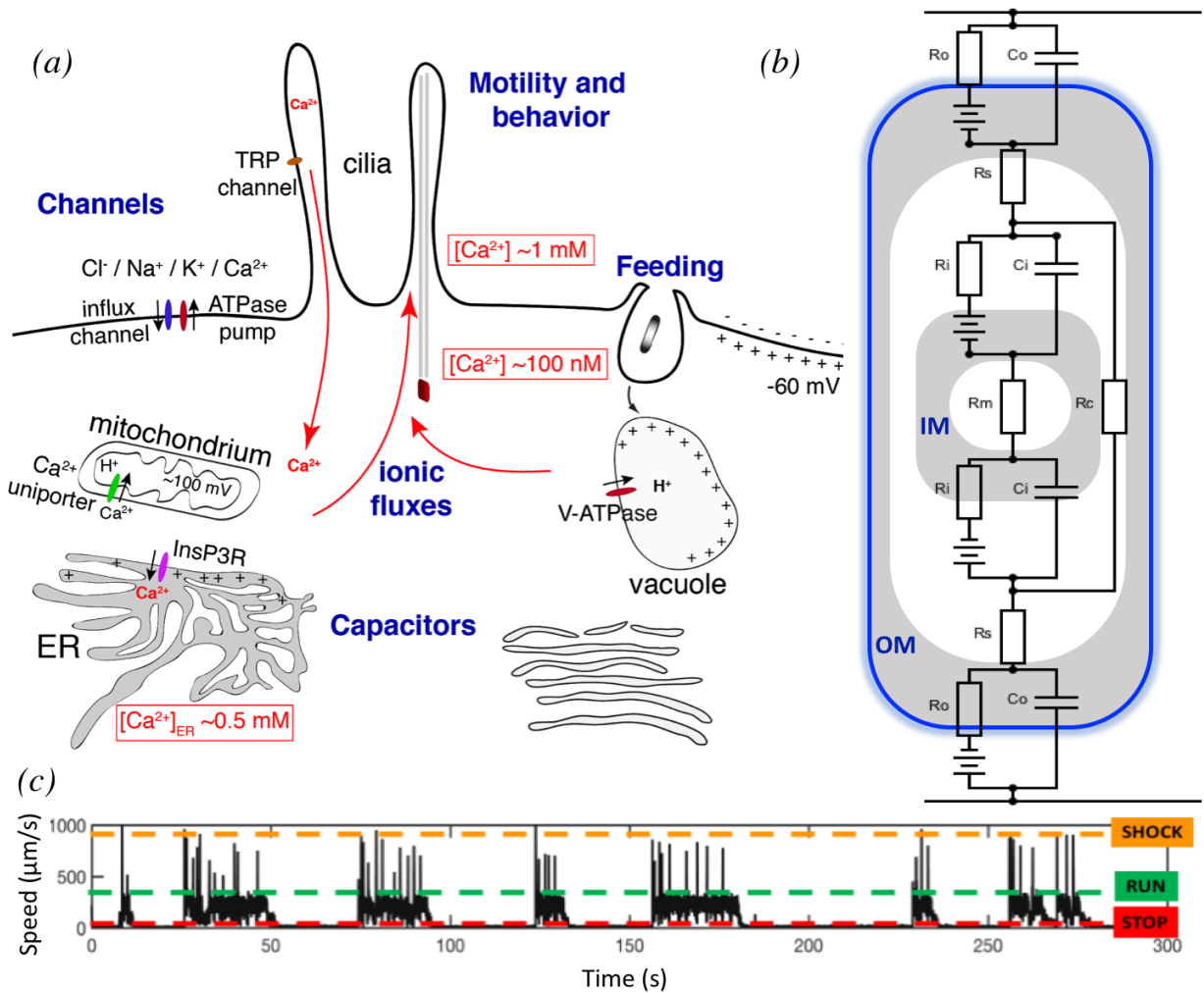

*Figure 2. Eukaryotic membranes control dynamic intracellular signalling.*
a) Massive diversification of lipids and ion channels creates compartments and circuits within a flexible and excitable endomembrane system. b) The mitochondrion has a distinctive two-layer topology with unique capacitative properties (IM - inner membrane, OM - outer membrane, Ci, Co are IM, OM capacitances, Ri, Ro, and Rm are IM, OM, and matrix resistances, Rc is a 'shunt' resistance [205] arising from cristae extending across the intermembrane space). c) Excitable signalling in cilia manifests as fast transitions or escape responses, as observed here in an octoflagellate protist [206] which switches between three distinct behavioural states (termed run, stop, and shock).

The capacity for subtle control of ciliary shape and movement underpins the exclusively eukaryotic trait of helical klinotaxis. Multiflagellate algae modulate the parameters of helical trajectories to realign with gradients in a graded, intensity-dependent manner – during *Chlamydomonas* phototaxis, photons incident on the eyespot induce a transient membrane depolarization [156] which alters the flagellar beat pattern [225]. Steering arises when the two flagella respond differently to this signal [226,227].

In contrast, during *E. coli* run and tumble, multiple flagella rotate in the same sense during swimming, but unbundling (when one or more motors rotate in the opposite sense) is stochastic. During stochastic chemotaxis, cells can only control the propensity for directional switching but not the new swimming direction. Some prokaryotes (e.g *Sinorhizobium meliloti*, *Rhodobacter sphaeroides*) can also control motor speed, e.g. [228–230].

Eukaryotes with multiple cilia can also have fine control over the movement sequence, or gait. This higher-level control requires the precise positioning of centrioles and cilia [231]. Multiciliary coordination can invoke hydrodynamic as well as intracellular means – including electrical signalling [124] or basal-body coupling [232]. Ciliary control dictates the walking gaits of *Euplotes* [233], as well as diverse swimming gaits in algal flagellates, including the breaststroke, trot, and gallop. Distinguished cilia on the same cell may be found alternately in sequences of excitation or quiescence [234].

Due to their larger size and active coordination, cilia can generate faster propulsion and flows. Collectively they can be organised across scales [226,235] for specific tasks. In ciliates and other filter feeders, cilia differentiate into oral cilia for feeding, and somatic cilia for swimming. Bioelectrical signalling controls mouth ciliature for complex feeding behaviours, as well as the ciliary beat direction to direct feeding flows [151].

Cilia arose not only to support eukaryotic motility but also sensing [236,237]. The continuity of the membrane around the cilium (and other eukaryotic appendages) is integral to these dual functions. In addition to the various sensory receptors localised to the cilium (discussed above), cilia and other eukaryotic appendages are sustained out of equilibrium by ATP-dependent mechanochemical cycles, which produce unique dynamical signatures [238,239]. Macroscale flux cycles in the phase space of movement patterns (broken detailed balance) have been observed in cilia [238], in hair cell bundles [240], in the motor patterns of intact swimmers [206] and in reconstituted cytoskeletal networks [240,241]. These exhibit active fluctuations that are orders of magnitude higher than thermal noise, and may further enhance sensory perception, particularly to mechanical stimuli and shear flows.

Besides cilia and pseudopodia, cytoskeletal polarization and patterning led to further, highly specialised forms of movement [242]. These involve dynamic mobilization of cellular protrusions for prey capture or manipulation (see Section 1E). Haptophyte algae have haptonema – a slender microtubular structure distinct from cilia, which coils at high speed to detect prey and other mechanical cues [243]. In *Vorticella* and *Zoothamnium*, contractile proteins in the stalk organelle (spasmoneme) can undergo extremely rapid mechanochemical contraction-extension cycles, fuelled by calcium [244,245]. The axopods of Heliozoa which are involved in feeding, contain bundles of microtubules, undergo rapid shortening by microtubule catastrophe [246,247]. In many predatory ciliates, ciliary bundles or cirri form specialised feeding structures which undergo dramatic shape changes to internalise prey.

### C. Origins of mitochondria and endomembranes as intracellular capacitors

The eukaryotic cell is distinguished from prokaryotic cells by a complex endomembrane topology [248–250]. The endomembrane system contains several charged compartments (Figure 2a) – multiple membranous structures, including the ER, the vacuole and mitochondria, often with closely-stacked lamellae (e.g. ER, plastids). This sophisticated structural organisation evolved during eukaryogenesis, and is critical to eukaryotic excitability.

One important structural innovation was the compartmentalisation of chemiosmotic ATP production. In eukaryotes, ATP synthesis occurs in the mitochondria, commonly during aerobic respiration [251]. In contrast, prokaryotes use their plasma membrane for ATP synthesis where ATP synthase complexes utilize the transmembrane electrochemical gradient of protons or $Na^+$ ions [252]. During eukaryogenesis, the original archaeal plasma membrane $A_0A_1$ ATP synthase [252,253] became internalized, evolving into the vacuolar H+-ATPase (ATP-driven proton pump) which acidifies vacuoles [254]. Free from the burden of chemiosmotic ATP production, the eukaryotic plasma membrane can now assume novel signalling roles.

Another important aspect of eukaryotic endomembrane organisation is the presence of several charged compartments with distinct ionic composition. These distinct compartments function as closed cellular capacitors [255,256]. The compartments are separated by membrane layers with low conductivity that form a physical barrier between the conductive internal and external fluid. These cellular capacitors actively release and replenish charges, gated by channels and pumps, which alter potential differences across membranes. In neurons, the speed of charge propagation from a synapse is inversely proportional to the specific capacitance ($C_m$, capacitance per unit area of membrane) of the membrane. For the plasma membrane of animal cells, this is estimated to be ~1 $\mu F/cm^2$ [257].

The ER reversibly stores and releases charge in the form of $Ca^{2+}$ ions. The tunable capacitance of the ER depends on the flow of external ions into the cell by capacitative $Ca^{2+}$ entry [258]. The acidified vacuole is positively charged relative to the cytosol, maintaining a vacuolar capacitance. Mitochondria and chloroplasts with their double membranes have unique capacitative properties (Figure 2b) [205,259].

We propose that in eukaryotes, the presence of multiple circuits consisting of these capacitors and their gating machineries represents a novel form of information storage and parallel processing not seen in prokaryotes. The organisation of the eukaryotic endomembrane system has three important functional consequences for cellular capacitance. First, the network of thin membranes creates a large surface area for charge storage and high capacitance – bilayer membranes are typically only 5 nm thick. For parallel plates (membanes) of area A separated by thickness d, and dielectric constant $\epsilon$, the capacitance is $C = \epsilon_0 A/d$. For a spherical capacitor of radius r, this is modified to $C = 4\pi\epsilon_0 r^2/d$. The capacitance of compressible membranes is also sensitive to mechanical compression, for a 5 nm thick membrane, a 1 nm decrease in thickness can significantly increase capacitance [260].

Second, membrane topology, comprising nested or closely-apposed membranes, greatly influences charge distribution. Where multiple membranes are stacked in parallel, resistances add reciprocally, while capacitances add linearly. The placement of different capacitors in a cell influences charge redistribution, particularly during dynamic phenomena such as motility and feeding. Membrane-bound organelles can be as close as 10 nm from the plasma membrane [261]. This physical proximity further ensures that coordinated signalling, or cross-talk, can occur near-synchronously across the different compartments. Changes in capacitance were measured in phagocytosing macrophages [261,262].

The third feature of the system is its ability to create and sustain nonlinear cycles of charging and discharging – a form of rapid bioelectric signalling. We illustrate this with a basic bilayer membrane (resistance R, capacitance C), here, the voltage-current relationship is given by $V = IR(1 - e^{-t/\tau})$, where $\tau = RC$ is a time constant for capacitative charging and discharging. This is typically on the order of 100 ns, but can be up to microseconds for larger cells [263]. These currents propagate throughout the cell [264], introducing temporal delays and thereby controlling the timing of signalling events, as has been demonstrated in nerve cells [265,266].

We conclude that during eukaryogenesis, the evolution of compartmentalized capacitors significantly increased the degrees of freedom available for intracellular electrical signaling, making critical contributions to eukaryotic excitability and behaviour. This critical function of the complex eukaryotic endomembrane system as a master regulator of behaviour and physiology extends beyond their bioenergetic or metabolic advantages. By analogy with electronics, eukaryotic cells constitute a complex and dynamic network of coupled resistors and interleaved capacitors as charge sources or sinks, that are associated with multiple time constants. Collectively, these circuits and motifs function as timers, frequency filters, tuners, whence complex behaviours can ensue.

### D. Flexible and excitable membranes

Except when covered by a cell wall, as are some fungal and plant cells, eukaryotic cells are morphable and undergo shape changes not seen in prokaryotes. One of the key steps of eukaryogenesis was the loss of the rigid glycoprotein cell wall of the archaea-derived host cell [267]. In archaea, the "*rigid structure and extremely tight adhesion or interdigitation of the glycoprotein cell walls [...] represent an invincible obstacle.*" [252].

The flexible plasma membrane in eukaryotes was a prerequisite for the evolution of total cell fusion, engulfment and membrane dynamics. The endomembrane system is suggested to have evolved rapidly, so that the LECA may have possessed a basic membrane-trafficking system of ER, endosomes, and stacked Golgi [268]. This was supported by a complex machinery of membrane-sculpting proteins (e.g. SNAREs, small GTPases, ESCRT complex, septins, coatomer) some of it with an origin in the Asgard archaea [269,270]. For example, according to sequence reconstructions, the LECA may have had as many as 23 Rab GTPases [271]. In the absence of a protective cell wall, eukaryotes may have become more prone to localised injury and membrane rupture. This could have contributed to the evolution of membrane repair and recycling mechanisms and excitable signalling [149].

Besides its flexibility, the plasma membrane harbours the greatest diversity of channels and receptors in eukaryotes, and – with the inclusion of the highly deformable ciliary membrane – became the most important surface for excitability. Eukaryotic membranes not only have diverse pumps and channels (see above) but also have thousands of distinct lipid species [272] compared to only hundreds in prokaryotic membranes. The same tendency for amphipathic lipids to self-assemble into layers occurs sub-cellularly in eukaryotes to form multiple boundaries which delimit individual organelles. Lipid diversity also encodes organellar identity (different compartments made up of different lipid species), thus preventing them from coalescing. This diversity may have indirectly contributed to maintaining distinct capacitative identities for electrical signalling in organelles.

Enrichment by eukaryotic sterols (particularly cholesterol) and sphingolipids, which form lipid rafts, are a signature eukaryotic trait [273]. Small changes in lipid structure can have dramatic consequences on membrane thermodynamics [274]. Raft clustering and sorting is thought to precede pinching and budding of vesicles, and are critical for membrane and vesicular trafficking. The composition of plasma membranes resides at a critical point [275], which has been proposed to reduce the energetic costs of membrane compartmentalisation and reconfiguration. Sterols are also major regulators of the activity of voltage-gated ion channels, which in turn function to alter the capacitance and conductance of eukaryotic membranes [248,276] (see previous section). A diverse compositional repertoire further ensures that leakages currents can be matched exactly.

Membrane shape depends on a complex interplay of proteins and lipids, and is highly sensitive to the heterogeneous distribution of lipids [277], which promotes the formation of bends and curvatures [278]. Excitability also depends on membrane fluidity [279]. Eukaryotic sterols have a profound effect on fluidity, channel permeability, and thickness [280]. Increased membrane fluidity was directly associated with changes in swimming speed in *Tetrahymena* [281]. In nerve membranes, anaesthetics are suggested to alter excitability by altering membrane fluidity [282].

Distortion of a fluid bilayer also provides a unique force-sensing mechanism as a form of membrane mechanosensitivity [283,284]. Shear forces and tension can activate transmembrane mechanosensory channels in a number of organisms, triggering action potentials [285]. Fluid shear has been shown to activate G-proteins [286], as well as increase membrane fluidity in endothelial cells [287] and the light-producing dinoflagellate *Lingulodinium polyedrum* (Alveolata) [288]. In another dinoflagellate *Pyrocystis lunula*, the critical threshold for bioluminescence was estimated to be ~0.1 μN, about thrice the shear stress on the cell wall [289]. Meanwhile the force required to activate rapid behavioural switches (shock response) in a small Prasinophyte alga can be under 10 pN [206]. Mechanical stimuli must do work to open transduction channels [290]. The amount of work required is a measure of sensitivity, which is much higher in eukaryotes (Table 2). In bullfrog hair bundles, $Ca^{2+}$ can shift the single-channel activation force by ~3 pN (equivalent to 1-2 $k_BT$) [291].

Eukaryotic membranes are capable of significant topological reorganisation, notably during wound-repair, phagocytosis, and cell-cell fusion. Plasma membranes reversibly deform during cell migration or feeding, to sustain non-energy-minimising structures such as slender protrusions [292]. The constant spatiotemporal remodelling and turnover of membranes is a eukaryotic trait. Vigorous membrane turnover is observed in some species of acanthamoeba, at an estimated complete turnover rate of several times per hour [293]. Membrane reorganisation, even fusion, can occur between predatory suctorians and their prey [294].

## E. Increase in cell size

If one were allowed to speak of a stereotypical eukaryotic cell, it would be about an order of magnitude larger than a stereotypical prokaryote. Cell size in single-celled eukaryotes ranges from 1 μm [295] to several cm (some macroalgae, such as *Acetabularia*), while prokaryotes are typically 1-10 μm, and can be diffraction-limited, the smallest mycoplasmas are ~0.2 μm [296]). Single-celled eukaryotes are found in most extant phyla. The largest ciliate has ~$10^6$ times the volume of the smallest eukaryote.

Cell size imposes fundamental limits on its ability to generate as well as control self-movement. We highlight two physical consequences of increasing size at the border between prokaryotes and eukaryotes. The first concerns the usefulness of self-movement for migration, while the second, the usefulness of self-generated flows for active feeding or nutrient sensing.

Virtually all single-celled organisms reside at low Reynolds numbers (Table 2), where viscosity dominates inertia [297]. At these scales there is no inertial coasting, so organisms stop instantaneously when they stop propelling themselves. Self-movement underlies navigation, active feeding, mating and many forms of excitability (Section 1, Figure 1). However, due to thermal noise, cells below a critical size can gain no absolute benefit from self-propulsion, regardless of the locomotion mechanism.

To see why, first consider an immotile cell of radius *a* immersed in a fluid of viscosity *μ*. It will be subject to Brownian motion, subject to both translational and rotational diffusion. The translational diffusion constant is given by the Stokes-Einstein law $D_0 = k_B T/(6\pi\mu a)$ and has units of length²/time, where $k_B T$ ($\approx 4.11 \times 10^{-21}$ J at room temperature) is the thermal energy at temperature T. Meanwhile rotational diffusion $D_r = k_B T/(8\pi\mu a^3)$ (units of radians²/time), gives rise to a timescale $\tau = 1/2D_r$ over which the orientation of the cell will be 'reset'.

An actively motile cell will have a higher effective translational diffusion $D_a$, which depends on the detailed movement strategy, or repertoire [11,15,18]. For run-and-tumble (random reorientations after tumble), $D_a \approx v^2\tau/3$. For prokaryotes, a typical run speed is $v = 10\mu m/s$ and maximum free-flight time $\tau$ can be limited by $D_r$ above. The benefit of active swimming to passive diffusion is quantified by $D_a/D_0 \sim v^2 a^4$, which improves rapidly with increasing size and speed. Dusenbery [298] used this expression to estimate that below a critical radius $a_c \approx 0.64\mu m$, active swimming is no longer useful. This may explain the paucity of motile bacteria with radius less than $a_c$. This motility divide is also present within the green-algal order Mamiellophyceae [299], which contains some of the smallest free-living eukaryotes: the non-motile *Ostreococcus* (~1 μm), and the motile *Micromonas* (~2 μm).

There is thus a dramatic advantage to having even a *slightly* larger size. Rotational diffusion $D_r$ determines how far a cell can move before a noise-driven trajectory reorientation [300]. For a prokaryote with $a = 1\mu m$, $D_r \approx 0.16$ rad²/s, this distance is at most $v\tau = v/2D_r \approx 3\mu m$, whereas for a eukaryote with $a = 10\mu m$, $D_r \approx 10^{-4}$ rad²/s, this is negligible. Active swimming enables eukaryotes to achieve a several fold-increase in sensory integration time and self-control over directional persistence (Table 2). Eukaryotes can therefore afford to move at higher speeds in order to achieve the same level of signal discrimination (more in 2F).

The second of our case studies focuses on how cell size influences solute transport. For a perfect spherical adsorber of radius *a*, the number of particles (present at ambient concentration $C_0$) per unit time, is given by $J = 4\pi D C_0 a$. Meanwhile an organism's metabolic needs scale much faster, according to volume, or $a^3$. The energy cost per unit volume of cell required to double the nutrient flux at the cell surface is estimated to be $\sim \mu D^2/a^4$ [301].

This means that while diffusion alone is sufficient to ensure prokaryotes are well-stirred, eukaryotes must have acquired alternate means of advecting flows. Cilia and flagella are often used to create large-scale flows (up to mm/s) for filter feeding, or for replenishing the local nutrient concentration around cells [302,303]. For a cell with length L, moving at speed U, the importance of advection over diffusion is captured by the Péclet number $Pe = UL/D$ [302,304], the ratio between advective (L/U) and diffusive timescales (L²/D). For a 1 μm prokaryote with a typical speed of 10 μm/s and $D \approx 10^{-5} cm^2/s$ (small molecules), $Pe = 0.01$, so no physiological amount of movement can improve diffusive uptake. Meanwhile eukaryotes readily achieve *Pe* numbers of order unity or above (Figure 3), so that active stirring is beneficial for nutrient redistribution (mass transfer), and to enable scanning through larger volumes of fluid [305].

While there are multiplicitous routes to increasing size and complexity, overcoming the diffusive bottleneck by active transport may have been a key step in eukaryogenesis, and multicellularity. Eukaryotic signature motilities (dynamic cytoskeleton) also enabled nutrient redistribution inside large cells. Increasing size presents new opportunities for creation of specialised organelles, such as mitochondria, feeding apparatuses, and sensors.

Importantly, most prokaryotes cannot steer deterministically in bulk fluid, nor is it advantageous to do so due to excessive rotational diffusion. Exceptions include a vibrioid bacterium [48], and magnetotatic bacteria including the bilophotrichous *Magnetococcus marinus [306]*, and the multicellular magnetotactic prokaryotes (~10 μm) [307]. The latter (also called MMPs) comprises multiple undifferentiated cells whose flagella exhibit some degree of coordination to achieve fast swimming and directional switching, aided by the Earth's magnetic field.

### F. New strategies for sensing and computation

All motile organisms larger than the critical size limit $a_c$ can access new strategies for sampling and sensing their environment beyond simple diffusion. Prokaryotes and eukaryotes use fundamentally distinct motility machinery, and have evolved distinct sensing mechanisms (with rare exceptions). Cells are remarkable sensors, able to monitor weak signals with a precision approaching physical limits [308]. Examples include chemoreception in bacteria [309,310], and in sperm [56], photon counting in vision [311], and frequency discrimination in the human ear [312]. It is often physical rather than biological processes which set the limits of cellular sensing and response at the microscale.

Below we discuss scaling laws which unequivocally constrain sensory fidelity at the microscale. These determine how the sensory signal-to-noise ratio depends upon the navigation strategy and the nature of the stimulus, through parameters such as cell size, swimming speed and sensory integration time. Here, we focus on limits imposed by different strategies for chemosensing and photosensing, and discuss the implications of these limits for cell evolution and the origins of eukaryotic excitability.

Thermal noise is ever-present, arising from molecular encounters (between receptors and ligands), photons, noise in chemical reactions, gene expression etc. Signal detection reduces to stochastic counting of molecules of chemoattractant, or photons (Figure 3A). For simplicity, we assume that gradients decay exponentially with a length scale $\ell_d$. Typically, for chemicals we can have $\ell_d \approx 0.1$ cm [298]. For light, the length scale will depend more strongly on the medium. We will use the same $\ell_d \approx 0.1$ cm for simplicity.

For chemosensing, molecules diffuse independently. So the number n of molecules bound by cellular receptors will be Poissonian, $n = 4\pi DC_0 \tau a$, where $D$ is the diffusion constant, $C_0$ the concentration, $\tau$ is the available integration or measurement time, and $a$ is the cell radius. During spatial sensing of a chemical, cells compare concentrations at two locations separated by $\Delta x = 2a$. Here the signal is $S = (J_+ - J_-)\tau = 6\pi DC_0 a^2 \tau / \ell_d$, where $J_\pm$ are the diffusive currents to two halves of a stationary sphere [298]. For a single measurement, the noise is $N = \sqrt{n}$ (Poissonian), with signal-to-noise ratio $SNR_s = S/N \sim a^{3/2} \tau^{1/2} D^{1/2} C_0^{1/2} / \ell_d$. If instead a motile cell moving at speed *v* performs temporal sensing, $S = C_0 \Delta x / \ell_d$, this time $\Delta x = v\tau$ (in the regime $\Delta x \ll \ell_d$). Here, the measurement error is reduced by taking multiple measurements over a time $\tau$, so that $N = C_0 (2\pi DC_0 a\tau)^{-1/2}$ [301], and $SNR_t = S/N \sim va^{1/2}\tau^{3/2}D^{1/2}C_0^{1/2}/\ell_d$. Similarly, during photosensing, the photon count $n = IAf\tau$ depends on *I*, the light intensity (photons per unit area per time), the area of the sensor *A*, the fractional absorbance *f* (<<1), and the integration time $\tau$. For spatial comparisons, $N = \sqrt{n}$, so that $SNR_s \sim (IAf\tau)^{1/2}(2a/\ell_d)$. Meanwhile for temporal comparisons, $SNR_t \sim (IAf\tau)^{1/2} v\tau/\ell_d$.

In all cases, the SNR depends strongly on $\tau$. For most prokaryotes this is severely limited by rotational diffusion. For light and chemical signals, the relative SNR between the two strategies scales similarly (with different prefactors): $SNR_s/SNR_t \sim a/v\tau$. Similar expressions can be derived for other cues, such as temperature, or for variations upon this sensory strategy [313]. Thus large, slowly moving cells will sense more effectively by spatial comparison, as in large amoeboid eukaryotes which crawl on substrates [26].

In contrast, small fast-moving cells use temporal sensing [314]. Below a certain size (~1 μm) cells become severely limited by rotational diffusion, and cannot maintain their orientation for long-enough to steer deterministically toward gradients. Such cells must adopt stochastic random walks. From $\sqrt{2D_r\tau}$ we estimate that in 1s the expected mean squared angular displacement is about 30° (Table 2). Only cells that are large enough can access the third strategy – helical klinotaxis. Here cells rotating around an axis can maintain stable helical swimming. Periodic stroke patterns produce superhelical trajectories over long times. Cells detect gradients perpendicular to the helix axis, and gradually steer toward gradients by adjusting the alignment between the swimming direction and gradient (error correction). Receptor organelles and sensors are localised to specific regions of the cell to further increase signal discrimination during active movement. For angular rotation speed $\omega$, we require $\omega \gg D_r$ which sets a minimum cell size for helical klinotaxis: e.g. a cell which rotates once per second about its axis must have a diameter of least 6 µm to achieve $\omega \approx 10 D_r$. Thus, klinotaxis is not physically accessible for most prokaryotes. Here the SNR for chemotaxis depends strongly on the signal gradient, and scales as $\sim R^2 T$, where R and T are the helix radius and period respectively [315,316].

Eukaryotic cells are in general several times larger than the wavelength of visible light and can use the cell body to focus light. For example, an eye-less strain of *Chlamydomonas* [317] or multicellular *Dictyostelium* slugs [318,319] can focus and align with directional light. In prokaryotes this mechanism was only described in cyanobacteria, which operate at the physical limit of focusing [49,320].

In reality, even spatial sensing strategies have a temporal component. Cells measure over a range of timescales, bound by a minimum reaction time, and a maximum memory recall. This leads to band-pass signal processing, as has been shown in some bacteria [321]. Since eukaryotes can access a wider range of timescales, they can respond to a wider range of signals. To achieve the same SNR, a larger size can compensate for faster movement, or a shorter reaction time. The entropic and energetic costs of performing cellular computations may have eventually promoted multicellularity or colonial-living, rather than building bigger and more complex single cells. In both pro- and eukaryotes, cell-cell communication can indeed enhance population-level sensing [322,323].

In summary, eukaryotes unlocked novel strategies for improving the sensory SNR. This principally results from an increase in size, polarization of sensors, and improved ability to control the swimming direction. The resulting behavioural stratification which partitions the eukaryotes from the prokaryotes can be visualised on a 2D plot of cell size and speed (Figure 3c).

| Physical parameter | Scaling | (Typical) prokaryote | (Typical) eukaryote | Comments |
|---|---|---|---|---|
| **Reynolds number** | Re = $\rho UL/\mu$ | Re = $10^{-5}$-$10^{-3}$<br><br>(L = 1-10 µm, U = 10 - 100 µm/s) | Re = $10^{-3}$-1<br><br>(L = 10-1000 µm, U = 10 - 1000 µm/s) | Unicellular eukaryotes and prokaryotes are viscosity dominated |
| **Péclet number** (fix D = $10^3$ µm$^2$/s for small molecule) | Pe = UL/D | Pe = $10^{-2}$-1<br><br>(L, U as above) | Pe = 1-$10^3$<br><br>(L, U as above) | Prokaryotes are diffusion-limited and cannot reach* Pe > 1 |
| **Diffusion constant (passive)** | $D_0 = k_B T/(6\pi\mu a)$ | $D_0$ = 0.22 µm$^2$/s<br><br>(a=1 µm) | $D_0$ = 0.02 µm$^2$/s<br><br>(a=10 µm) | Eukaryotes 10x less susceptible to linear diffusion |
| **Rotational diffusion (passive)** | $D_{rot} = k_B T/(8\pi\mu a^3)$ | $D_{rot}$ = 0.16 rad$^2$/s<br><br>(a=1 µm) | $D_{rot}$ = 1.6×$10^{-4}$ rad$^2$/s<br><br>(a=10 µm) | Eukaryotes 1000x less susceptible to rotational diffusion |
| **Effective diffusion (active, depends on motility strategy)** | $D_a = v^2 \tau/3$<br><br>($\tau$ = a free-flight time, v = speed of runs) | $D_a \sim$ 133 µm$^2$/s<br><br>(v ~ 20 µm/s, $\tau$ ~ 1s) | $D_a \sim$ 3.3×$10^{-2}$ cm$^2$/s<br><br>(v ~ 100 µm/s, $\tau$ ~ 10 s) | Empirical:<br><br>*E.coli:* [324] D ~ 10-100 µm$^2$/s<br><br>*C. reinhardtii:* [325] D ~ $10^{-3}$ cm$^2$/s<br><br>*P. caudatum:* [326] D ~ $10^{-2}$ cm$^2$/s |
| **Relevant for stochastic navigation** | Expected angular deviation in a time $\tau$<br>$\theta_{rms} = \sqrt{(2D_{rot}\tau)}$ | $\theta_{rms}$ = 32°<br><br>(for $\tau$ = 1s) | $\theta_{rms}$ = 1°<br><br>(for $\tau$ = 1s) | Prokaryotes are severely limited by rotational diffusion (cannot steer) |

| **Relevant for spatial sensing** | $\frac{SNR_{spatial}}{SNR_{temporal}} \sim \frac{a}{v\tau}$ | Too small* | Useful strategy for large and slow-moving cells | Common strategy for amoeboid eukaryotes |
|---|---|---|---|---|
| **Relevant for helical klinotaxis** | $SNR_{klinotaxis} \sim R^2/\omega$ | Too small* | Useful strategy to reorient toward vectorial cues | Common strategy for free-swimming eukaryotes |
| **Sensitivity to mechanical stimuli (membrane tension)** | Ratio of channel opening and closing probabilities $\frac{P_o}{P_c} = \exp(-\frac{\Delta E}{k_B T})$ ΔE = work done to open channel (~sensitivity) | Bacterial MscS & MscL channels (osmotic nanovalves): [283] 5~10 mN/m | Hair cells, single channel gating stiffness: [290] ~1 mN/m Piezo1: [327] ~1.4 mN/m | 1 $k_B T$ is 4 nm$^2$ change in area under tension of 1mN/m Lytic tension of pure lipid bilayer ~20 mN/m |

*Table 2. Summary of key biophysical scaling relationships.*
See main text for references and for *rare prokaryotic examples which defy the general trend. (We have assumed throughout: $k_B T = 4.11 \times 10^{-21}$ J = 4.11 pN nm, and $\mu/\rho = 10^{-2}$ cm$^2$ s$^{-1}$ is the kinematic viscosity of water at room temperature.)

## 3. Concluding remarks

With increasing size and structural complexity, eukaryotes developed novel strategies for environmental exploration. Different eukaryotic lineages are characterised by unique combinations of feeding modalities, control pathways, and organization of motility appendages. Eukaryote-signature capabilities are most clearly manifest in emergent mechanisms of motility control, particularly of membrane-bound appendages such as cilia. Eukaryotes, thus endowed, became capable of programmed responses to complex stimuli in a more finely-tuned manner. In this article, we have identified specific yet universal changes in cellular and membrane architecture which we propose were crucial for such organisms to overcome many of the challenges associated with life and survival.

Though our list may not be exhaustive, it offers a refreshing perspective for tracing the possible scenarios of eukaryogenesis, which emphasizes the need to consider signalling and behaviour at the whole-cell level. As prokaryotes evolved colonial-living (e.g. biofilms) to overcome the bottlenecks encountered at the microscale in terms of sensing and fluid-interactions, single-celled eukaryotes instead were able to attain a level of behavioural sophistication unmatched in single-celled prokaryotes. Given the strong scaling of size with rotational diffusion, we find that even a modest increase in size may have been sufficient to enable greater self-control and movement persistence (Figure 3), and therefore significant improvement in signal discrimination and decision-making. Such capabilities likely coevolved with cellular and structural innovations.

We suggest that interfacial excitability – the capacity for small potential differences to feedback and suddenly amplify – is critical for the novel functionality of eukaryotic membranes. Although the question of why prokaryotic membranes do not appear to be capable of propagating reproducible, millisecond action potentials remains open, the answer is likely to lie with the massive diversification of proteins and lipids in eukaryotic membranes. We propose that the compositional complexification of eukaryotic membranes conferred excitability to eukaryotes. This diversification enhanced the precision with which control can be exerted over the thermodynamic state of the membrane, its capacitance, fluidity, as well as the identity and magnitude of transmembrane ion fluxes. In this regime, even small changes in physical parameters can feedback and amplify nonlinearly.

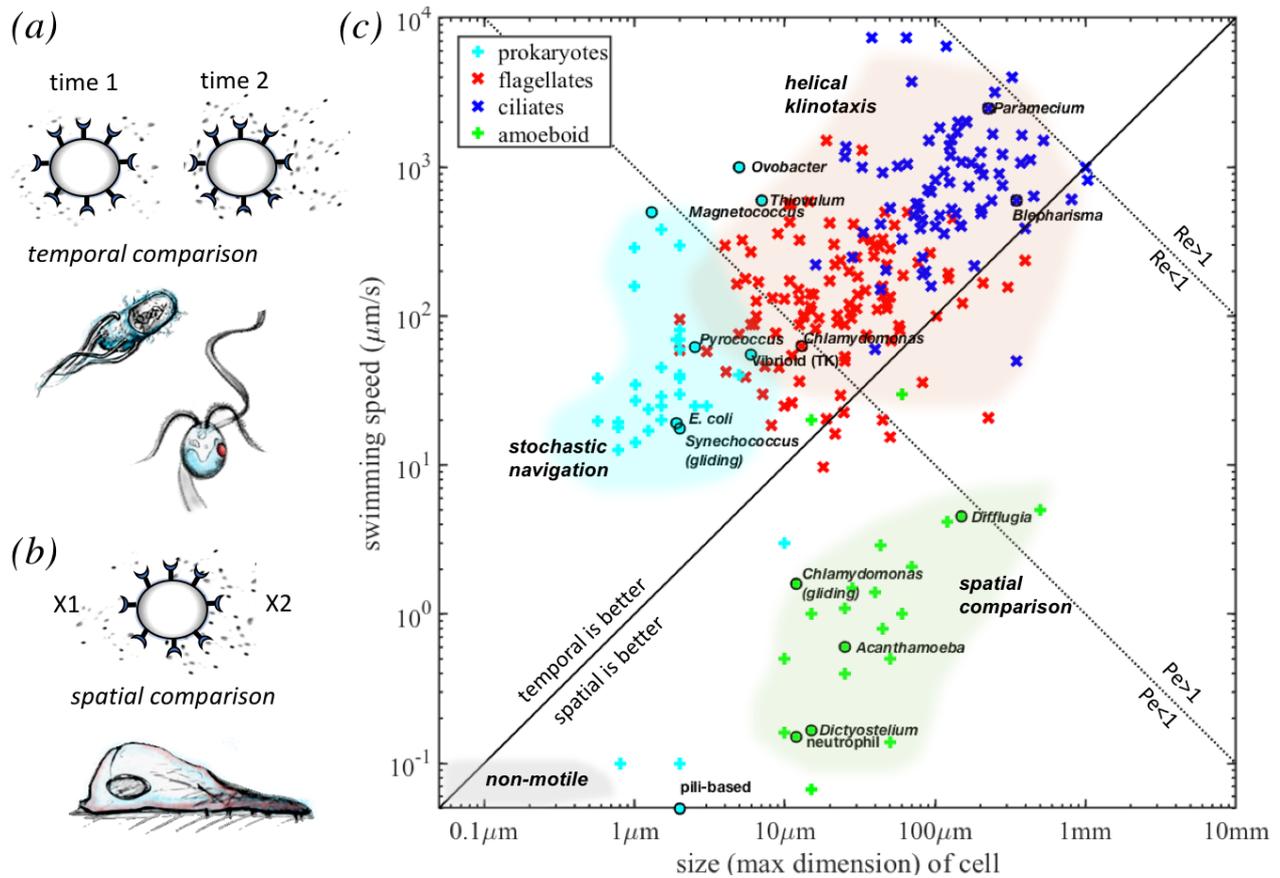

*Figure 3. Eukaryotes unlocked new biophysical regimes.*
a) Temporal sensing strategies typically involve comparing the signal at two slightly different times - this can be stochastic and rotational-diffusion limited (as in many prokaryotes), or involve self-steering (as in many ciliates and flagellates). b) Spatial sensing strategies involve comparing the signal at two different positions at the same time (as in most amoeboid cells). (c) These distinctions create a prokaryote-eukaryote divide with respect to behaviour, as visualised here in a phase space of organism size-versus-speed (log-log scale). Three phase boundaries partition this space (see also Table 2), namely the Reynolds number (*Re*), Péclet number (*Pe*), and the relative advantage of temporal vs spatial sensing (plotted here for an integration time $\tau = 1$ s). Data points are collated from the literature, where some species of particular interest are highlighted. *(In particular see: amoeba [328], flagellates and ciliates [329], marine bacteria [330]. The full dataset will be available as a spreadsheet upon publication.)*

In eukaryotes, the integration of fast sensory and excitable motility elements may thus have unlocked new physical regimes of memory and information processing in cells (for example, spatiotemporal navigation requires significant computational bit depth). The capacity to precisely control and generate action potentials, leading to fast escape responses or behavioural transitions, may be a strategy for self-protection from accidental damage. Single-cell neuronal computation [331] may thus have had ancient roots tracing back to ancestral eukaryotes which enacted distributed control over motile appendages [234].

The acute "environmental sensibility" of excitable membrane interfaces was probably a key trait of ancestral eukaryotes. Excitability enabled eukaryotes to move, sense, react, feed and finally to fulfil the Biblical injunction, to *multiply* (paraphrasing Mills, Peterson and Spiegelman [332]), in radically new ways. It would be hard to escape the conclusion that excitability had a major contribution to the success of eukaryotes and enabled them to conquer all but the most extreme habitats on the planet. What we hope we could also show is that the divide separating eukaryotes from prokaryotes is as wide for excitability and behaviour as for most other cellular features. Future models of eukaryogenesis ought to account for the origins of excitability to paint a fuller picture of this momentous transition in evolution.

**Acknowledgements**

KYW gratefully acknowledges funding from the European Research Council (ERC) under the European Union's Horizon 2020 research and innovation programme under grant agreement No 853560: "EvoMotion" - *Moving around without a brain: Evolution of basal cognition in single-celled organisms*. GJ would like to thank the Leverhulme Trust for funding (RPG-2018-392).

158. Fujiu K, Nakayama Y, Yanagisawa A, Sokabe M, Yoshimura K. 2009 Chlamydomonas CAV2 encodes a voltage- dependent calcium channel required for the flagellar waveform conversion. *Curr. Biol.* **19**, 133–139.
159. Umbach JA. 1981 *PH and Membrane Excitability in Paramecium Caudatum*.
160. Dunlap K. 1977 Localization of calcium channels in Paramecium caudatum. *J. Physiol.* **271**, 119–133.
161. Echevarria ML, Wolfe GV, Taylor AR. 2016 Feast or flee: bioelectrical regulation of feeding and predator evasion behaviors in the planktonic alveolate Favella sp. (Spirotrichia). *J. Exp. Biol.* **219**, 445–456.
162. Lueken W, Ricci N, Krüppel T. 1996 Rhythmic spontaneous depolarizations determine a slow-and-fast rhythm in walking of the marine hypotrich Euplotes vannus. *European Journal of Protistology*. **32**, 47–54. (doi:10.1016/s0932-4739(96)80038-1)
163. Kunita I, Kuroda S, Ohki K, Nakagaki T. 2014 Attempts to retreat from a dead-ended long capillary by backward swimming in Paramecium. *Front. Microbiol.* **5**, 270.
164. Wood DC. 1988 Habituation in Stentor: produced by mechanoreceptor channel modification. *J. Neurosci.* **8**, 2254–2258.
165. Jennings HS. 1899 Studies on Reactions to Stimuli in Unicellular Organisms. III Reactions to Localized Stimuli in Spirostomum and Stentor. *The American Naturalist*. **33**, 373–389. (doi:10.1086/277256)
166. Dexter JP, Prabakaran S, Gunawardena J. 2019 A Complex Hierarchy of Avoidance Behaviors in a Single-Cell Eukaryote. *Curr. Biol.* **29**, 4323–4329.e2.
167. Yang C-Y, Bialecka-Fornal M, Weatherwax C, Larkin JW, Prindle A, Liu J, Garcia-Ojalvo J, Süel GM. 2020 Encoding Membrane-Potential-Based Memory within a Microbial Community. *Cell Syst* **10**, 417–423.e3.
168. Grishanin RN, Bibikov SI, Altschuler IM, Kaulen AD, Kazimirchuk SB, Armitage JP, Skulachev VP. 1996 delta psi-mediated signalling in the bacteriorhodopsin-dependent photoresponse. *J. Bacteriol.* **178**, 3008–3014.
169. Williams TA, Cox CJ, Foster PG, Szöllősi GJ, Embley TM. 2020 Phylogenomics provides robust support for a two-domains tree of life. *Nat Ecol Evol* **4**, 138–147.
170. Zachar I, Szathmáry E. 2017 Breath-giving cooperation: critical review of origin of mitochondria hypotheses : Major unanswered questions point to the importance of early ecology. *Biol. Direct* **12**, 19.
171. López-García P, Moreira D. 2020 The Syntrophy hypothesis for the origin of eukaryotes revisited. *Nat Microbiol* **5**, 655–667.
172. Schäffer DE, Iyer LM, Burroughs AM, Aravind L. 2020 Functional Innovation in the Evolution of the Calcium-Dependent System of the Eukaryotic Endoplasmic Reticulum. *Front. Genet.* **11**, 34.
173. Jackson AP *et al.* 2016 Kinetoplastid Phylogenomics Reveals the Evolutionary Innovations Associated with the Origins of Parasitism. *Curr. Biol.* **26**, 161–172.
174. Stathopulos PB, Ikura M. 2017 Store operated calcium entry: From concept to structural mechanisms. *Cell Calcium* **63**, 3–7.
175. Prakriya M, Feske S, Gwack Y, Srikanth S, Rao A, Hogan PG. 2006 Orai1 is an essential pore subunit of the CRAC channel. *Nature* **443**, 230–233.
176. Bick AG, Calvo SE, Mootha VK. 2012 Evolutionary diversity of the mitochondrial calcium uniporter. *Science* **336**, 886.
177. Alzayady KJ, Sebé-Pedrós A, Chandrasekhar R, Wang L, Ruiz-Trillo I, Yule DI. 2015 Tracing the Evolutionary History of Inositol, 1, 4, 5-Trisphosphate Receptor: Insights from Analyses of Capsaspora owczarzaki Ca2+ Release Channel Orthologs. *Mol. Biol. Evol.* **32**, 2236–2253.
178. Plattner H, Verkhratsky A. 2015 The ancient roots of calcium signalling evolutionary tree. *Cell Calcium* **57**, 123–132.
179. Cai X, Clapham DE. 2012 Ancestral Ca2+ signaling machinery in early animal and fungal evolution. *Mol. Biol. Evol.* **29**, 91–100.